\documentclass[]{article}
\def\half{{1\over 2}}
\def\ben{\begin{equation}}
\def\een{\end{equation}}
\def\Dslash{\slash \negthinspace \negthinspace \negthinspace \negthinspace D}

\def \p{\partial}
\input amssym.def
\input amssym.tex
\begin{document}
\def \mathbb{\Bbb} 
\def\cramp{\medmuskip = 2mu plus 1mu minus 2mu}
\def\cramper{\medmuskip = 2mu plus 1mu minus 2mu}
\def\crampest{\medmuskip = 1mu plus 1mu minus 1mu}
\def\uncramp{\medmuskip = 4mu plus 2mu minus 4mu}
\def\half{{\textstyle{1\over2}}}
\let\a=\alpha \let\b=\beta \let\g=\gamma \let\d=\delta \let\e=\epsilon
\let\z=\zeta \let\h=\eta \let\q=\theta \let\i=\iota \let\k=\kappa
\let\l=\lambda \let\m=\mu \let\n=\nu \let\x=\xi \let\p=\pi \let\r=\rho
\let\s=\sigma \let\t=\tau \let\u=\upsilon \let\f=\phi \let\c=\chi \let\y=\psi
\let\w=\omega      \let\G=\Gamma \let\D=\Delta \let\Q=\Theta \let\L=\Lambda
\let\X=\Xi \let\P=\Pi \let\S=\Sigma \let\U=\Upsilon \let\F=\Phi \let\Y=\Psi
\let\C=\Chi \let\W=\Omega
\let\la=\label \let\ci=\cite \let\re=\ref
\let\se=\section \let\sse=\subsection \let\ssse=\subsubsection
\def\nn{\nonumber}
\let\fr=\frac \let\bl=\bigl \let\br=\bigr
\let\Br=\Bigr \let\Bl=\Bigl
\let\bm=\bibitem
\let\na=\nabla
\let\pa=\partial \let\ov=\overline
\newcommand{\be}{\begin{equation}}
\newcommand{\ee}{\end{equation}}
\def\ba{\begin{array}}
\def\ea{\end{array}}
\def\ft#1#2{{\textstyle{\frac{\scriptstyle #1}{\scriptstyle #2}}}}
\def\fft#1#2{\frac{#1}{#2}}
\def\del{\partial}
\def\vp{\varphi}
\def\sst#1{{\scriptscriptstyle #1}}
\def\oneone{\rlap 1\mkern4mu{\rm l}}
\def\td{\tilde}
\def\wtd{\widetilde}
\def\ie{\rm i.e.\ }
\def\dalemb#1#2{{\vbox{\hrule height .#2pt
        \hbox{\vrule width.#2pt height#1pt \kern#1pt
                \vrule width.#2pt}
        \hrule height.#2pt}}}
\def\square{\mathord{\dalemb{6.8}{7}\hbox{\hskip1pt}}}
\newcommand{\ho}[1]{$\, ^{#1}$}
\newcommand{\hoch}[1]{$\, ^{#1}$}
\newcommand{\bea}{\begin{eqnarray}}
\newcommand{\eea}{\end{eqnarray}}
\newcommand{\ra}{\rightarrow}
\newcommand{\lra}{\longrightarrow}
\newcommand{\Lra}{\Leftrightarrow}
\newcommand{\ap}{\alpha^\prime}
\newcommand{\bp}{\tilde \beta^\prime}
\newcommand{\tr}{{\rm tr} }
\newcommand{\Tr}{{\rm Tr} }
\def\0{{\sst{(0)}}}
\def\1{{\sst{(1)}}}
\def\2{{\sst{(2)}}}
\def\3{{\sst{(3)}}}
\def\4{{\sst{(4)}}}
\def\5{{\sst{(5)}}}
\def\6{{\sst{(6)}}}
\def\7{{\sst{(7)}}}
\def\8{{\sst{(8)}}}
\def\n{{\sst{(n)}}}
\def\cA{{{\cal A}}}
\def\cB{{{\cal B}}}
\def\cF{{{\cal F}}}
\def\tV{\widetilde V}
\def\tW{\widetilde W}
\def\tH{\widetilde H}
\def\tE{\widetilde E}
\def\tF{\widetilde F}
\def\tA{\widetilde A}
\def\im{{{\rm i}}}
\def\tY{{{\wtd Y}}}
\def\ep{{\epsilon}}
\def\vep{{\varepsilon}}
\def\R{\rlap{\rm I}\mkern3mu{\rm R}}
\def\bD{{{\bar D}}}

\def\R{\rlap{\rm I}\mkern3mu{\rm R}}
\def\bD{{{\bar D}}}
\def\R{{{\mathbb R}}}
\def\C{{{\mathbb C}}}
\def\H{{{\mathbb H}}}
\def\CP{{{\mathbb C}{\mathbb P}}}
\def\RP{{{\mathbb R}{\mathbb P}}}
\def\HP{{{\mathbb H}{\mathbb P}}}
\def\A8{{{\mathbb A}_8}}
\def\Z{{{\mathbb Z}}}
\def\bA{{{\mathbb A}}}
\def\bB{{{\mathbb B}}}
\def\bC{{{\mathbb C}}}
\def\bD{{{\mathbb D}}}
\def\bE{{{\mathbb E}}}
\def\bZ{{{\mathbb Z}}}
\def\Re{{{\frak{Re}}}}
\def\Im{{{\frak{Im}}}}
\def\cosec{{\,\hbox{cosec}\,}}
\def\Gm{{\Gamma_{\!\! -}}}
\def\Gp{{\Gamma_{\!\! +}}}
\def\stan{{standard }}
\def\nonstan{{supernumerary }}
\def\gcd{{\rm gcd}}
\def\kdel#1{{\fft{\del}{\del#1}}}

\newcount\hour \newcount\minute
\hour=\time  \divide \hour by 60
\minute=\time
\loop \ifnum \minute > 59 \advance \minute by -60 \repeat
\def\nowtwelve{\ifnum \hour<13 \number\hour:
                      \ifnum \minute<10 0\fi
                      \number\minute
                      \ifnum \hour<12 \ A.M.\else \ P.M.\fi
         \else \advance \hour by -12 \number\hour:
                      \ifnum \minute<10 0\fi
                      \number\minute \ P.M.\fi}
\def\nowtwentyfour{\ifnum \hour<10 0\fi
                \number\hour:
                \ifnum \minute<10 0\fi
                \number\minute}
\def\now{\nowtwelve}

\thispagestyle{empty}
\hfuzz=100pt
\title{Solitons and Black Holes in 4 and 5 Dimensions}
\author{
\\
Department of Applied Mathematics and Theortical Physics \\
\\  University of Cambridge ,
\\ Silver Street,,
\\ Cambridge CB3 9EW,
\\ U.K.
}

\maketitle

\begin{abstract}
Two lectures given in Paris in 1985. 
They were circulated as a preprint 
Solitons And Black Holes In Four-Dimensions, Five-Dimensions.
G.W. Gibbons (Cambridge U.) . PRINT-85-0958 (CAMBRIDGE), 
(Received Dec 1985). 14pp.
and appeared in print in
De Vega, H.J. ( Ed.), Sanchez, N. ( Ed.)
: Field Theory, Quantum Gravity and Strings*, 
46-59 and Preprint - GIBBONS, G.W. (REC.OCT.85) 14p.

I have scanned the original, reformatted and 
and corrected various typos.

\end{abstract}

\centerline{\today\,\,\now}

\tableofcontents

\section{ Introduction} 

This is the written version of two lectures given in Paris in February
1985.  Since the material as given has now appeared elsewhere [1,2]  I
have decided not to repeat the lectures verbatim but rather  to
comment on the general problem of solitons in gravity, in particular
on the importance or otherwise of spatial and spacetime topology
contrasting the situation in 4 and in 5 spacetime dimensions. My main
point will be that while there are many similarities with the
situation in Yang-Mills-Higgs theory there are significant
differences. In particular the apparently inevitable occurrence of
spacetime singularities and their conjectured shielding by event
horizons (Cosmic Censorship) means that one cannot assume that the
time evolution of initial data  is continuous. This substantially
alters ones views of the importance  of topology in the classical
theory. It is highly likely that the quantum theory - should it ·make
mathematical sense - will be similarly affected. The plan of the
article is as follows: in section 2 I will discuss some topological
aspects of the initial value problem. In section 3 I will describe why
I don't feel one can regard black holes  as solitons  except in
the extreme Reissner-Nordstrom case, and the relation of this to
supergravity. In section 4 I will contrast the situation with that in
5-dimensions and I will argue that the true analogue of magnetic
monopoles in Yang-Mills theory are the multi-Taub NUT solutions whose
importance for Kaluza-Klein theory was first  stressed by Gross, Perry
and Sorkin. Their relation to black holes  will also be described. In
section 5 I will describe a duality  conjecture analogous to that of
Olive \& Montonen in the Yang-Mills  case. 

\section{ Topology and the Initial Data} 
It is an attractive idea that the way to study solitons and  other
topological features in General Relativity is to start with an
initial data set $(\Sigma, g_{ij}, K_{ij}) $ where $\Sigma$  is a
3-dimensional manifold, $g_{ij}$ a  Riemannian metric and $K_{ij}$
the second fundamental form.  The metric and second fundamental form
should satisfy certain constraints and be  asymptotically flat. Indeed
one could imagine more than one asymptotic region, just as there is in
the Schwarzschild vacuum solution. The $k$ asymptotic regions may be
imagined to be compactified to give a compact manifold $\bar \Sigma$ ,
$\Sigma$  being diffeomorphic to $\bar \Sigma$  with $k$ points
removed.  There is no complete topological classification of  3
manifolds but it is known [3]  that for orientable manifolds  $\Sigma$
may be  expressed uniquely as the connected sum of a number of "prime
manifolds"  $\Sigma_i$  \ben {\bar \Sigma}= \Sigma_1 \# \Sigma _2
\dots \#  \Sigma _n  \een A complete list of prime manifolds is not
known  but it is known that for instance $S^2 \times S^1$  and
elliptic spaces $S^3 /\Gamma$  where $\Gamma$  is a suitable discrete
subgroup of $S0(4)$ with free action on $S^3$ are prime.  Initial data
satisfying the constraints  which are orientable are, according to
Schoen and Yau [4]  probably limited to a sum of $S^2 \times  S^1$'s
and elliptic spaces. 

The existence   of a unique factorization has led Witten [5] to argue
that there are  no solitons in 4-dimensional gravity because if there
were one would expect an anti-soliton 3-metric such that one  \ben S^3
= \Sigma_s \# {\bar {\Sigma _s}}   \een where $\Sigma _s$  is the
soliton 3-space topology and and  ${\bar {\Sigma _s}}$  that of the
anti-soliton.  If $\Sigma _s$   is prime this is ruled out by the
uniqueness. One  now seems to have a problem with CPT since (2)
implies that the  soliton anti-soliton pair cannot have the quantum
numbers of the vacuum. The way out of this particular difficulty would
seem to be that topology is not a "good quantum number". This seems
reasonable because it appears that any topologically non-trivial
initial data set must evolve to give spacetime singularities in its
future [6].

According to the widely believed but still as yet unproven Cosmic
Censorship Hypothesis [7] these singularities will be shielded inside
event horizons. Furthermore it is also widely believed that the final
state (in the classical theory) will consist of one or more time
independent black holes. These black holes will have the metric of the
Kerr solution.  The consequences of this are rather disappointing as
far as spatial topology is concerned. Suppose one started with for
instance one of Sorkin's non-orientable wormholes [8]. That is  ${\bar
  {\Sigma _s}}=P $  the non-orientable $S^2$  bundle over $S^1$ .  It
is not difficult to construct initial data with this topology [9].
This has a number of fascinating topological properties [3].  For
instance, topologically:  \ben P \# (S^2 \times S^1 ) \equiv P \#  P
\een  which one might interpret as saying that  two non-orientable
wormholes  could turn into a non-orientable wormhole and a
conventional orientable wormhole. All of this however will be
invisible from infinity since presumably each, or maybe both, will be
surrounded by event horizons and the fact that they are topologically
non-trivial will play no role in the exterior dynamics; The final
black hole solution will be a Schwarzschild or Kerr metric and no hint
of the interior topology will show up in that. 

Very much the same applies to the significance of the $\theta$
-vacuum structure of the initial data.  One might view the
configuration space $Q$ for gravity as the space of Riemannian metrics
on ${\bar {\Sigma _s}} $ factored by the set of diffeomorphisms $ {\rm
  Diff}^*( {\bar {\Sigma _s}} $ ) having a point on   ${\bar {\Sigma
    _s}} $  (the point at infinity) and its tangent space
invariant. If $ {\rm Diff}^*( {\bar {\Sigma _s}} $ ) is not connected
the configuration space $Q$  will not be simply connected and $\theta$
-vacuum analogous to those in Yang-Mills theory are possible [10].  A
particular instance of this is the beautiful work of Sorkin and
Friedman [11] on spin ${1 \over 2}$  from gravity. Because $Q$ is not
simply connected a rotation of the spacetime  relative to infinity may
result in one  moving around a closed loop in $Q$  which is not
homotopic  to the constant path. The quantum wave function could in
principle change sign under such a rotation.  As an example consider
as they do  $ {\rm Diff}^+( {\bar {\Sigma _s}} $ )to be $ S^3/\Gamma $
where $\Gamma$  is the 8 element group consisting of the  unit
quaternions and their negatives together with $\pm 1$.  It is quite
easy to construct time symmetric initial data corresponding to this
space. The resulting space can be thought of as containing  7 black
holes suitably identified [9].  Despite the exotic topology it seems
rather likely that the end result  will be will be just one large
black hole. Again there will be no sign in the external metric of the
initial exotic topology. 

Finally as a final argument against the significance of  3-space
topology let me remind the reader of the well known theorem of
Serini, Einstein, Pauli and Lichnerowicz which I like to paraphrase as
"No solitons without horizons". The theorem states that there are no
regular globally static solutions of the vacuum Einstein equations
other than the flat one. The argument depends on the fact that if
$g_{00}=-V^2 $  with $ V >0$  on $\Sigma$  and $ V \rightarrow  1$ at
infinity the field equations  imply that  \ben \nabla _i \nabla ^i
V=0\,,\een  where $\nabla _i$ is covariant differentiation with
respect to the spatial metric $g_{ij}$ .  The maximum principle
immediately shows that $V=1$  The remaining  field equation now reads
\ben R_{ij}=0 \een where $R_{ij}$ is the Ricci tensor of $g_{ij}$
which in 3-dimensions shows that $g_{ij}$  and hence the 4-dimensional
metric must be flat.

\section{ The Black Hole as Soliton} 
 
The remarks in section 2 have been intended to convince the reader of
the importance of the 4-dimensional dynamics of the theory as opposed
to that of 3-dimensional~ initial data. This does not mean that one
can necessarily regard black holes as solitons. Far from it. They have
no fixed mass or angular momentum even in the classical theory. Indeed
the non-decreasing property of the event horizon area is anything but
solitonic. The situation is even worse in the quantum theory since we
know from the work of Hawking [12] that black holes are unstable
against thermal evaporation. We are still ignorant of the final
outcome of this process which may not be calculable in Einstein theory
but may require  a consistent quantum theory of gravity. A plausible
guess is that the hole simply disappears in a puff of radiation. If
this is true the black hole should be regarded in the quantum theory
as an unstable "intermediate state", rather than a stable
particle-like state. 

The exception to this would be if the hole carried a "central"
charge. By central I mean completely conserved and not carried by any
of the fundamental fields of the theory. For example in N=2 ungauged
extended supergravity. [13] there is a Maxwell field. The fields of
the N=2 supergravity multiplets are the graviton, the photon and the
gravitino. These are all electrically neutral with respect to the
Maxwell field - that is why the theory is "ungauged". It is quite
possible for black holes to carry this charge - essentially because
the lines of flux are "trapped in the topology" as people used to say
in the days of "Geometrodynamics". The metric of such holes (if
nonrotating) is that of Reissner and Nordstrom. It is parameterized by
the mass M and charge Q. Because of the duality, invariance of the
theory of any magnetic charge may be rotated to zero by a suitable
duality rotation. The singularity is clothed by an event horizon if
\ben M \ge{ |Q| \over \kappa}\qquad \een where $\kappa ^2 = 4 \pi G$
and $G$  is Newton's constant.  I have described in more detail
elsewhere [1,2] how one may view (6) as a Bogomolny type inequality
[see also 14,15,16,17]. The electric charge $Q$ is truly central in
the sense of the super symmetry algebra and the inequality in (6) is
saturated by extreme black holes which are "supersymmetric" in that
they possess "Killing spinors". There exist a whole family of
multiblack hole metrics [17] satisfying (6). These are the
PapapetrouMajumdar metrics [18] which are included in the general
class of Israel-Wilson metrics [19]. Tod [20] has shown that the
Israel-Wilson metrics exhaust all the metrics with Killing spinors in
N=2 supergravity. It has been known for some time that the throat of
the extreme ReissnerNordstrom metric has the geometry of the
Robinson-Bertotti solution,  i.e. the product metric on $S^2 \times
{AdS}_2 $  where $ {AdS} _ 2$  is 2-dimensional anti-de Sitter space.
The Robinson-Bertotti metric shares with flat  space the property of
being maximally supersymmetric - i.e. of having  the largest possible
number of Killing spinors; Thus the extreme Reissner-Nordstrom metrics
spatially interpolate between the 2 possible "vacua" of .N=2 ungauged
supergravity. The possible relevance of this remark for spontaneous
compactification is intriguing. For the present let me remark that
this is typically soliton-like behaviour. 

Since the charge is central it cannot be lost during Hawking
evaporation and so a hole with an initial charge must settle down to
the lowest mass state with that charge. This is the extreme (zero
temperature) state. Thus extreme Reissner-Nordstrom holes seem to
behave just like solitons. The hole with the opposite charge is
clearly the anti-soliton and it seems extremely plausible that a
soliton-anti-soliton might completely annihilate one another. They
cannot do this classically if Cosmic Censorship holds since by
Hawking's area theorem the final event horizon must have non-vanishing
area but the resultant Schwarzschild black hole can then evaporate
thermally .  The main way in which the extreme holes differ from
solitons is that there seems to be no way of fixing their mass or
charge - i.e. no quantization rule.  (6)  Since the extreme holes
(which need not all have the same mass) , can remain in equilibrium it
is reasonable to consider departures from equilibrium
perturbatively. To lowest order they should move  on geodesics on a
suitable "moduli space", that is to lowest order the parameters
specifying the solution should change slowly. This is the same
approximation as has been used successfully in Yang-Mills theory
[21,22].  In the present case the Papapetrou~Majumdar solution
(representing N black holes) is specified by giving the positions of
N points in ${\Bbb R}^3 $.  In principal the points could coincide
though I will argue in a short while that this doesn't happen. If the
holes, having equal masses, were identical one would factor by the
action of the permutation group $S_N$  on the N positions. Thus we
know the moduli space.  The metric is not known. However if one makes
the approximation that one hole is very much smaller than all  the
others, one can anticipate that the motion of the small hole  in the
field of the others should be given by the standard equation for a
charged geodesic  (with charge =   mass $\times \kappa $).  In the
slow motion limit this does indeed give non-relativistic  geodesic
motion in the metric  \ben ds^2 = U^3 d{\bf x} ^2  \een where  \ben
U=1 + \sum _{i=1}^{i=N-1} {GM_i \over |{\bf x} -{\bf x}_i| }  \een

This metric is complete on ${\Bbb R}^3 - \{{\bf x}_i \} $.
In this approximation the holes . 
would take an infinite time to merge or coalesce.

The quantum scattering of extreme holes could be studied in the
non-relativistic limit by looking at the Schr\"odinger equation on the
moduli space. This would presumably correspond to the scalar Laplacian
with respect to the metric on the moduli space, though it is also
possible that potential terms might appear due to one loop effects.
In the case that the holes all had equal mass one should divide out by
the permutation group. The moduli space would have fundamental group
$S_N$.  The wave function could in principle then be even or odd under
permutation. Thus one could imagine "fermionic" black holes! This is
the analogue of the effect of Sorkin and Friedman I described above.
It is possible to find extreme black holes in the N=4 ungauged
extended supergravity theory as well [23]. They should also probably
be thought of as solitons. Like the extreme holes in N=2 they also
have no natural mass quantization - at least as far as classical or
semi-classical considerations are concerned. To get a satisfactory
quantization rule one seems forced to turn to Kaluza-Klein theory.

\section{Solitons in 5-dimensions} 
Much of the discussion about the relevance of topology in section  2
could be repeated here with 4 replacing 3. The details of the
topological discussion would differ and we certainly don't have
detailed singularity theorems and black hole uniqueness theorems in
higher dimensions - indeed we know very little about black holes in
higher dimensions. However in higher dimensions gravity is even more
attractive (having a force inversely as distance to the power of the
dimension of spacetime minus 2) than in 4-dimensions. In 5-dimensions
it depends on distance in the same way as the repulsive centrifugal
force (which is inversely as distance cubed in all dimensions). In
higher dimensions  it rises even more rapidly than the centrifugal
repulsion. This would seem to make gravitational collapse and
spacetime singularities even more likely in higher dimensions.
However there is an important difference. We are no longer obliged, nor
would we wish, to confine ourselves to initial data which are
asymptotically Euclidean. If we do so the argument that the vanishing
of the Ricci tensor implies that the 4-space is flat still goes
through according to Schoen and Yau's Positive Action Theorem [25]. If
we  don't require that the 4-metric be asymptotically Euclidean there
are many complete Ricci flat 4-metrics, including one - that on the K3
surface - which is compact. Any gravitational instanton will give a
static 5-metric with no horizons. Note that if we have no horizon we
are still forced to have V = 1, that is the metric must be a product
on ${\Bbb R}  \times  M$ , where $M$ is the 4-manifold. In the older
language the  space time would be said to be "ultrastatic".  Not all
of these objects will be classically stable. The stability will be
governed by spectrum of the Lichnerowicz Laplacian acting on symmetric
tensors on M. If M has a self-dual metric this is known to be positive
and hence the corresponding static lump will be classically stable. If
M has a metric which is not self-dual the spectrum is not   likely to
have a positive spectrum and the corresponding lump will be
unstable. Examples of this are the Euclidean Schwarzschild solution
[26] and the "Taub-Bolt solution" [27]. 
 
The evolution of these objects in the full non-linear theory is
unclear.  The Euclidean Schwarzschild solution has the same
asymptotics as the flat metric on ${\Bbb R} ^3 \times S^1$ and  so
presumably it loses energy to gravitational radiation and attempts to
settle down to the flat metric but it can't do this without forming
some sort of singularity since this would involve a spatial topology
change. It seems likely that a black hole will be formed but this is
not known. The same remarks apply to the Taub-Bolt metric which
presumably tries to settle down to the  Taub-NUT metric.  Again black
hole formation seems likely.  It is possible that these black holes
appear regular when viewed from a 4-dimensional stand-point in which
case they should be included with those described in [24] and
[28]. The Hawking effect may then cause these black holes to evolve to
the flat or the Taub-NUT solution. 

The boundary conditions of interest for Kaluza-Klein theory is that
the metric be what has been called in this context asymptotically flat
i.e.  that it approach the flat product metric on  ${\Bbb R} \times
S^1  $  at infinity  or that it be asymptotically locally flat.  The
typical example of the latter is the  self-dual Taub-NUT metric, or
multi-Taub-NUT with N centres.  The topology at infinity in this case
is ${\Bbb R} \times B_N$  where  $B_n$  is the $S^ 1$  bundle over
$S^2$  with Hopf invariant N - i.e. the lens space  $L(N, 1)$ .

Gross, Perry and Sorkin [29] have pointed out that the Taub-NUT
solution plays the role of a magnetic monopole in Kaluza-Klein
theory. Perry and myself [30] have shown that the monopole moment P of
any asymptotically locally flat solution should satisfy the Bogomolny
type inequality  \ben {|P| \over 2 \kappa } \le M   \een

with equality in the supersymmetric self-dual case.
 It is interesting to note that the gravitational instanton solution of 
Atiyah \& Hitchin [22] is self-dual but has a {\sl negative}  mass. 
This is presumably because it has the topology at infinity of 
${\Bbb R} \times (S^3 /\Gamma ) $  where $\Gamma$ 
is the binary dihedral group. The crucial point here is whether 
or not suitable solutions of the Witten equation exist. 

The multi Taub-NUT solutions are specified by giving  N non-coincident
points in ${\Bbb R} ^3$  . Permutating the points gives the same
metric  so the moduli space is the well known configuration space  $ (
({\Bbb R}^3 ) ^N - \Delta) / S_N$  where  $\Delta$ is the set of
points in $(({\Bbb R})^3 )^N$  where two or more points coincide  and
$S_N$  is as before the permutation group on N symbols.  The metric
on the moduli space is under study.  Again the quantum mechanics
offers the possibility of multivalued wave functions though whether
these monopoles can really be thought of as fermions remains at
present unclear.  An important property of the Taub-NUT solutions is
that the magnetic charge  $P$  satisfies the Dirac quantization
condition:  \ben eP = 2 \pi  \een  where $e$  is the basic unit of
charge in Kaluza-Klein theory.  This in turn implies that (using the
equality in (9))  the mass $M$ is quantized:  \ben M = { 1 \over 4 \pi
  \kappa e}  \een

Given their stability and the quantization of the mass and magnetic
charge it seems reasonable to regard the Taub-NUT solutions as
representing solitons though this does require, as in section 2, that
some of the topological numbers associated with the object are not
conserved. In the present case two such numbers are of interest. The
Hirzebruch signature and the Euler number. The multiple monopole  has
non-vanishing Hirzebruch signature. Roughly it corresponds to magnetic
charge. Since this can be read off from the asymptotic boundary
conditions one might expect this to be conserved. The Euler number is
a different matter however. This cannot be determined from infinity
and given the likely occurrence of singularities there seems to be no
good reason for it to be conserved. Another argument, due to Hawking,
is that the Euclidean action in General Relativity is not scale
invariant. This means that it may cost arbitrarily little action to
pass from one topological configuration to another. This is unlike the
case in Yang-Mills theory in 4-dimensions where the action is scale
invariant and 	 typically topologically different configurations
differ by an amount $8 \pi ^2 \over g^2 $  where $g$ is  the coupling
constant. If one does accept them as solitons one sees a number of
striking resemblances with the massive modes of the Kaluza-Klein
theory. This is the subject of the next section.

\section{ Pyrgon-Monopole duality} 
 
The physical content of the 5-dimensional Kaluza-Klein theory when
viewed from the point of view of 4-dimensions 

\begin{itemize}
\item 1) A set of {\sl massless} states, the graviton, graviphoton and dilaton 

\item 2) A tower of {\sl massive}  states of spin 0, 1 and 2 each
 with mass $m$  and charge $e$ given by 
\ben
m= n { |e| \over 2 \kappa}  \een 
where $ n = 1,2,3,\dots $ . 
\end{itemize}

At the linearized level all the massive states are trivially stable.
When one takes into account interactions one might expect  the higher
mass statesto decay into  lower mass states but a charged state cannot
decay into a neutral state.  Thus the lowest mass states, $n=1$ ,
should be absolutely stable except against annihilation wiith their
antiparticle states. These stable lowest mass states have been called
Pyrgons [31].  Thus the perturbative physical Hilbert space  consists
of massless states, pyrgons and antipyrgons.  In a supersymmetric
theory the pyrgons fit into massive supermultiplets with central
charge. In N=8 for example the relation (12) corresponds to the
maximal central charge allowed.  This is necessary to avoid states
with spin greater than 2.

Now the G-P-S monopoles possess in the N=8 supergravity model of
Cremmer  [32] the maximum permitted number of Killing spinors and
hence supersymmetries. As shown in [30] they fit into supermultiplets
when the zero modes are taken into account. There is a rather close
analogy, indeed one is tempted to say a duality, between the monopoles
of Kaluza-Klein theory and the pyrgons. This suggested duality is
analogous to that which has been suggested in Yang-Mills theory
[33]. In the present case we suggest that there might exist in the
full quantum theory operators which create and annihilate monopole
states. In addition there will be operators which create the massless
states. If these satisfy an effective field theory it is essentially
unique -  it must be the original field theory of the pyrgons. This is
essentially because of the supermultiplets structure. Thus we have the
conjectured dualities:  \bea \nonumber {\rm monopole}
&\leftrightarrow&  {\rm pyrgon} \\ \nonumber {\rm massless \, fields}
&\leftrightarrow&{\rm massless \,   fields} \\ \nonumber{\rm
 antimonopole} &\leftrightarrow&   {\rm anti-pyrgon} \eea

It is difficult to see with present day techniques how such a
conjecture could be verified. In the Yang-Mills case some partial
evidence has come from a study of magnetic and electric dipole
moments. It has been verified that the gyromagnetic ratio of the
ordinary YangMills particles equals the gyroelectric ratio of the
monopoles plus fermionic zero-modes [34]. It is known that the
gyromagnetic ratios of the pyrgons are anomalous and equal unity,
rather than the Dirac value of 2 [35]. It would be interesting to
calculate the electric dipole moments of G-P-S monopoles with their
fermionic zero-modes.  Further insight into this conjectured duality
might come from a study of monopole-pyrgon interactions. A number of
authors [36] have pointed out that there is no "Callan-Rubakov" effect
[37] which would catalyze the decay of pyrgons. This is mos~ easily
seen from the fact that scalar modes on Taub-NUT are well defined and
using the covariantly constant spinor fields on Taub-NUT one can
obtain all solutions of the Dirac equation.  Thus if $\epsilon$  is
a covariantly constant spinor on  Taub-NUT and a solution of the
wave equation with energy $\phi_\omega$  then: 
$$
\psi_\pm = (\phi_\omega \pm {1 \over \omega} 
(\Dslash   \phi _\omega) ) \epsilon 
$$
are solutions of the Dirac equation with the same energy. 

A striking fact about the scalar modes on the Taub-NUT background is
that the massive scalar pyrgon wave equation separates in 2  different
coordinate systems.  One system is the standard radial variables in
which the metric is \ben ds^2 =(1 + {2 N \over \rho})^{-1} 4N^2 (d
\psi + \cos \theta d \phi)^2 + (1 + {2 N \over \rho}) ( d \rho ^2 +
\rho ^2 ( d \theta ^2 + \sin ^2 \theta d \phi ^2 ))  \een where $0\le
\psi \le 4 \pi$. Thus $8 \pi N= 2 \pi R_K$ , where  $R_K$   is the
radius of the Kaluza-Klein circle.  The scalar field has the form \ben
\phi_\omega = e^{-i\omega t} e^{i{n \over 2} \psi} \, { _{n\over
    2}Y_{lm}} (\theta ) e^{im\phi} f_n(\rho)  \een where $ { _{n\over
    2}Y_{lm}} (\theta ) e^{im\phi}  $   is a spin weighted spherical
harmonic and where $f_n(\rho)$   1s a non-relativistic Coulomb wave
function with angular momentum  $l$  but where the Coulomb potential
is energy dependent, i.e. depends  upon $\omega$  , that is $f$
satisfies \ben { 1 \over \rho ^2 }  {d \over d \rho }(\rho ^2 {d f
  \over d \rho}) -{l(l+1) f \over \rho ^2 } +  ( 2N \omega ^2 -{n^2
  \over N} ) { f \over \rho } + (\omega ^2 -{n^2 \over 4N^2 } ) f=0
\een

There are no bound states, just scattering states. 
Since the radial equation (15) is a Coulomb one one 
might anticipate that scattering is 
better described using parabolic coordinates, defined by 
$$ \xi =  \rho (1 + \cos \theta ) \qquad \eta = \rho (1- \cos \theta)  
$$

This is in fact true. 
The wave equation also separates in the $t,\phi ,\xi ,\eta $
 coordinates. Using them one can give a simple description of the 
scattering. The classical orbits are especially simple
 being {\sl conic  sections} . 
They are, when projected into the 3-space spanned by 
$\rho, \theta \phi$,  the intersection of a cone centred at $\rho =O$
 with a plane, the intersection being a hyperbola in general. 

The existence of 2 different coordinate systems in which the wave
equation separates is often taken as the indication of hidden
symmetries and indeed of a "spectrum generating algebra".  The precise
nature of this algebra in the present case has not been worked out. It
is tempting to speculate that it may be related to  the known
existence of Kac-Moody algebras in Kaluza-Klein theory [38]. 

Another tempting speculation is that these ideas will find their full 
expression in string theory. Michael Green [39]
 has remarked that if one considers 10-dimensional string theory 
where   10-D of the spacelike dimensions form a torus, 
each of whose radii equals $R$ 
one obtains string states with masses satisfying 
\ben
({\rm mass}) ^2 = \sum ^\infty _{i=0} ( {M^2_i  \over R^2 } + { R^2 N^2_i \over
{\alpha ^\prime}^2 } ) + { 2 \over \alpha ^\prime} (N_0 + {\tilde N}_0 )  
\een

$N_0$  and ${\tilde N} _0)$  are occupation numbers 
for higher string states. 
The integers $\{M_i\}$ 
are Kaluza-Klein charges resulting from the periodicity in the 
10-D compact dimensions. The integers $\{N_i\}$  are topological 
charges associated with the number round the i'th compact dimension. 
Consider the limit
\ben
R \rightarrow 0\,\quad  {\rm and} \quad  {\alpha ^\prime \over R}\,\quad  {\rm constant} = \lambda  
\een

The resulting D-dimensional 
field theory has an infinite number of massive spin 
2 supermultiplets whose masses are determined by $\lambda$  
This theory is apparently identical to the 
theory obtained by starting with 10-dimensional N=2 
supergravity and compactifying on a hypertorus with 
(10-D) dimensions having finite radii $R = \lambda$ ,

Now set $D=5$ . 
The reduction of N=2 d=10 to 5 dimensions gives 
Cremmers 's N=8 D=5 model, with its pyrgon states. 
On the other hand from (17) we see that 
the states corresponding to zero Kaluza-Klein charge but 
non-vanishing topological winding numbers will survive in this limit. 
These presumably correspond to the magnetic monopole states.


\begin{thebibliography}{99}



\bibitem {[1]}  G.W. Gibbons in "Supersymmetry, Supergravity and Related Topics" ed. F. del Aguila, J.A. de Azcarraga and L.E. Ibanez, World Scientific 1985. 

\bibitem {[2]} G.W. Gibbons in "Non Linear phenomena in Physics" ed. F. Claro, Springer Proceedings in Physics \#3. Springer Verlag 1985. 


\bibitem {[3]} J. Hempel "Topology of 3-Manifolds", Princeton University Press 
(1976) . 


\bibitem {[4]} R. Schoen and S.T. Yau, Phys. Rev. Lett: 1457 (1979). 

\bibitem{[5]}  E. Witten, Cornrnun. Math. Phys. 100 197 (1985). 

\bibitem {[6]} D. Gannon, J. Math. Phys. 2364 (1975). 
D. Gannon, G.R.G. 7 219 (1976). 
C.W. Lee, Commun. Math. Phys. 51 157 (1976). 


\bibitem {[7]} R. Penrose, Ann. N.Y. Acad. Sci  224 125 (1973). 

\bibitem{[8]}  R. Sorkin, J. Phys. A10 717 (1977) 


\bibitem {[9]} G.W. Gibbons, unpublished. 

\bibitem {[10]} C.J. Isham, Phys. Lett. 106B 188 (1981). 
C.J. Isham, in "Quantum Structure of Space and Time", eds. C.J. Isham 
\& M.J. Duff. Cambridge University Press (1982). 

\bibitem {[11]} J. Friedman \& R. Sorkin, Phys. Rev. Lett. 44 1100 (1980); 
	see also B. Witt: Milwaukee preprint.  

\bibitem {[12]} S.W. Hawking, Nature (Lond.) 248 30 (1974). 
S.W. Hawking, Commun. Math. Phys. 43 199 (1975). 

\bibitem {[13]} S. Ferrara \& P. van Nieuwenhuizen, Phys. Rev. Lett. 
37 1669 (1976) . 

\bibitem{[14]}  G.W. Gibbons, in Heisenberg Memorial Symposium,
 ed. P. Breitenlohner and H.P. Durr, Springer Lecture Notes in Physics \#160. 

\bibitem {15]} G.W. Gibbons \& C.M. Hull, Phys. Letts. 109B 190 (1982). 

\bibitem {[16]} G.W. Gibbons, in Proc. 4th Silarg Symposium, ed. C. 
Aragone, World Scientific. 

\bibitem {[17]} J.B. Hartle and S.W. Hawking, Commun 
. Math. Phys. 26 87 (1982).

\bibitem{ [18]}  A. Papapetrou, Proc. Roy. Irish. Acad. A51 191 (1947). 
S.D. Majumdar, Phys. Rev. 72 390 (1947)

\bibitem {[19]} W. Israel and G.A. Wilson,~. Math. Phys. 13 865 (1972). 

\bibitem {[20]} P. Tod, Phys. Lett. B121 241 (1983). 	-- 

\bibitem {[21]} N. Manton, Phys. Rev 
N. Manton in "Monopoles in Quantum Field Theory" ed. 
N. Craigie, P. Goddard and W. Nahm, World Scientific (1982). 

\bibitem {[22]} M.F. Atiyah and N. Hitchin, Phys. Lett. 107A 21 (1985). 

\bibitem {[23]} G.W. Gibbons, Nucl. Phys. B207 337 (1982) 

\bibitem {[24]} W. Simon, G.R.G. 17 761 (1985). 

\bibitem {[25]} R. Schoen and S.T, Yau, Phys. Rev. Lett. 42 547 (1979). 

\bibitem {26]} D. Page, Phys. Rev. B. Allen, Phys. Rev. 030 1153 (1984). 

\bibitem {[27]} R.E. Young, Phys. Rev:-D28 2420 (1983). 

\bibitem {[28]} G.W. Gibbons and D. Wiltshire, Annals of Phys., in press.

\bibitem { [29]} D. Gross and M.J. Perry, Nucl. Phys. 8226 29 (1983). 
R. Sorkin, Phys. Rev. Lett. 51 87 (1983). 

\bibitem {[30]} G.W. Gibbons and M.J. Perry,
Nucl. Phys. B248 629 (1984). 

\bibitem{[31]} E.W. Kolb and R. Slansky, Phys. Lett. 135B 378 (1984). 

\bibitem {[32]} E. Cremmer  in "Superspace and Supergravity", 
ed. S.W. Hawking and S.W. Hawking \& M. Rocek, 
Cambridge University Press 1981. 

\bibitem {[33]} C. Montonen and D.l. Olive( Phys. Lett. 72B 117 (1977).

\bibitem {[34]} H. Osborn, Phys.Lett. 115B 226 (1982). -- 
    Bo,Yu. Hou, Phys. Lett.125B 389 (1983). 

\bibitem {[35]} A. Hoysoya et al., Phys. Lett. 134B (1984).


\bibitem {[36]} P.C. Nelson, Nucl. Phys. 238B 63s-(1984). 
     H. Ezawa and A. lwasaki, Phys. Lett. 138B 81 (1984). 
     M. Kobayashi and A. Sugamoto, Prog. Theor. Phys. 72 122 (1984).
      F.A. Bais and P. Batenburg, Nucl. Phys. B245 469 (1984). 

\bibitem {[37]}  V. Rubakov., Pisma Zh. Eksp. Teor. Fiz. 58 (1981); Nucl. 
	Phys. 203B 311 (1982). 	C.G. Callan, Phys. Rev. D25 2141 (1981). 

\bibitem {[38]} A. Salam and J. Strathdee, Annals of Phys. 141 316 (1982). 
L. Dolan and M.J. Duff, Phys. Rev. Lett. 52 (1984). 


\bibitem {[39]}  M. Green "The Status of Superstrings" 
undated Queen Mary College preprint. 




\end{thebibliography}
\end{document}